\documentclass[useAMS,usenatbib]{mn2e}

\usepackage[dvips]{graphicx}
\usepackage[]{txfonts}
\def\simless{\mathbin{\lower 3pt\hbox
{$\rlap{\raise 5pt\hbox{$\char'074$}}\mathchar"7218$}}}   
\def\simmore{\mathbin{\lower 3pt\hbox
{$\rlap{\raise 5pt\hbox{$\char'076$}}\mathchar"7218$}}}   
\newcommand{\be}{\begin{equation}}
\newcommand{\ee}{\end{equation}}
\newcommand{\ubs}{Swift J1644+57}
\newcommand{\ubsb}{Swift J2058+05}

\title[]{Afterglow model for the radio emission from the jetted tidal disruption candidate {\it Swift} J1644+57}
\author[B. D. Metzger, D.~Giannios, P.~Mimica]
{Brian D.~Metzger$^{1,3}$\thanks{E-mail: bmetzger@astro.princeton.edu},
  Dimitrios Giannios$^{1}$, Petar Mimica$^{2}$\\
$^{1}$Department of Astrophysical Sciences, Peyton Hall, Princeton
  University, Princeton, NJ 08544, USA\\
$^{2}$Departamento de Astronom´a y Astrof´sica, Universidad de Valencia, 46100, Burjassot, Spain\\
$^{3}$NASA Einstein Fellow\\}

\begin{document}
\date{Received / Accepted}
\pagerange{\pageref{firstpage}--\pageref{lastpage}} \pubyear{2011}

\maketitle

\label{firstpage}

\begin{abstract}

The recent transient event \ubs~has been interpreted as emission from a collimated relativistic jet, powered by the sudden onset of accretion onto a supermassive black hole following the tidal disruption of a star.  Here we model the radio$-$microwave emission as synchrotron radiation produced by the shock interaction between the jet and the gaseous circumnuclear medium (CNM).  At early times after the onset of the jet ($t \lesssim 5-10$ days) a reverse shock propagates through and decelerates the ejecta, while at later times the outflow approaches the Blandford-McKee self-similar evolution (possibly modified by additional late energy injection).  The achromatic break in the radio light curve of \ubs~is naturally explained as the transition between these phases.  We show that the temporal indices of the pre$-$ and post$-$break light curve are consistent with those predicted if the CNM has a wind-type radial density profile $n \propto r^{-2}$.  The observed synchrotron frequencies and self-absorbed flux constrain the fraction of the post-shock thermal energy in relativistic electrons $\epsilon_{e} \approx 0.03-0.1$; the CNM density at 10$^{18}$ cm $n_{18} \approx 1-10$ cm$^{-3}$; and the initial Lorentz factor $\Gamma_{\rm j} \approx 10-20$ and opening angle $\theta_{\rm j} \sim (0.3-1)\Gamma_{\rm j}^{-1} \sim 0.01-0.1$ of the jet.  Radio modeling thus provides {\it robust independent evidence for a narrowly collimated outflow}.  Extending our model to the future evolution of \ubs, we predict that the radio flux at low frequencies ($\nu \lesssim$ few GHz) will begin to brighten more rapidly once the characteristic frequency $\nu_{m}$ crosses below the radio band after it decreases below the self-absorption frequency on a timescale of months (indeed, such a transition may already have begun).  Our results demonstrate that relativistic outflows from tidal disruption events provide a unique probe of the conditions in distant, previously {\it inactive} galactic nuclei, complementing studies of normal AGN.

\end{abstract} 
  

\section{Introduction} 
\label{intro}

Supermassive black holes (SMBHs) are most easily studied when they accrete at high rates for extended periods of time, powering active galactic nuclei (AGN).  Although AGN provide important probes of the central regions of distant galaxies, this view is necessarily biased because periods of bright AGN activity are special epochs in the history of a galaxy, during which the substantial kinetic and photon luminosities from the SMBH substantially alter the distribution of gas and dust from their more typical ``quiescent'' state.  Because the majority of SMBHs are underfed, the conditions in most galactic nuclei are thus challenging to study most of the time.

The high energy transient {\it Swift} J164449.3+573451 (hereafter \ubs) was detected by the Burst Alert Telescope (BAT) on March 28, 2011 as a low SNR image trigger.  Subsequent imaging at radio, optical and, X-ray wavelengths localized the event to within $\lesssim 150$ pc of the center of a compact galaxy at redshift $z \simeq 0.353$ (\citealt{Berger+11}; \citealt{Fruchter+11}; \citealt{Levan+11b}).  The inferred R-band luminosity places an upper limit $\lesssim 10^{7}M_{\odot}$ on the mass of the SMBH in the galactic center, if the galaxy follows the \citet{Magorrian+98} relation (\citealt{Levan+11}; \citealt{Bloom+11}; \citealt{Burrows+11}; cf.~\citealt{Miller&Gultekin11}).  \citet{Quataert&Kasen11} point out that the SNR of the {\it Swift}/XRT data is not sufficient to constrain significant variability on $<$ 10 s timescales, in which case the mass of the central compact object could in principle be much lower, possibly consistent with a stellar core collapse event, as are associated with normal long-duration gamma-ray bursts (GRBs).  However, the coincidence of \ubs~with the galactic nucleus, combined with the lack of any previously known GRBs with similar luminosity or duration \citep{Levan+11}, suggest that \ubs~most likely originated from the rapid onset of accretion onto a SMBH \citep{Bloom+11}.  A high energy transient with similar properties, {\it Swift} J2058.4+0516 (hereafter \ubsb), was recently discovered by \citep{Cenko+11}, who argue that this event also resulted from rapid accretion onto a SMBH in a galaxy at redshift $z \simeq 1.18$.

Lack of evidence of prior radio or X-ray activity \citep{Levan+11} suggest that the rapid increase in the accretion rate responsible for \ubs~and \ubsb~cannot result from gas entering the sphere of influence of the SMBH (for instance from the destruction of a passing molecular cloud), since this would require a timescale $\sim r_{soi}/\sigma > 10^{4}$ yr to appreciably alter the accretion rate near the horizon, where $r_{soi}\sim 1$ pc is the radius of the sphere of influence and $\sigma \sim 100$ km s$^{-1}$ is a typical bulge velocity dispersion.  A more plausible source of such a rapid onset of accretion is the tidal disruption of a star by a SMBH (tidal disruption event, or ``TDE''; \citealt{Rees88}).  Analytic estimates and numerical calculations show that the process of disruption leaves a significant fraction of the shredded star gravitationally bound to the black hole (e.g.~\citealt{Rees88}; \citealt{Ayal+00}; \citealt{Lodato+09}).  The accretion of this stellar debris has long been predicted to power a thermal `flare' at optical, UV, and X-ray wavelengths that lasts for months to years after the merger (e.g.~\citealt{Ulmer99}; \citealt{Strubbe&Quataert09}; \citealt{Lodato&Rossi11}).  Until recently, this thermal signature was the only means known for detecting TDE candidates (e.g.~\citealt{Komossa&Greiner99}; \citealt{Donley+02,Gezari+08,Gezari+09,vanVelzen+10}).

Just a few months prior to the discovery of \ubs~, \citet{Giannios&Metzger11} (hereafter GM11) explored the observational consequences if a modest fraction of the accretion power following a TDE is used to accelerate a collimated jet to ultra-relativistic speeds.  Regardless of what ultimately powered the outflows responsible for \ubs~and \ubsb, if the nuclear region was indeed previously inactive then such short-lived ejections cannot propagate far from the SMBH before beginning to decelerate via their interaction with the dense gas of the circumnuclear medium (CNM).  This is in stark contrast to long-lived AGN jets, which are active for tens of millions of years, during which time the jet evacuates a cavity of kpc scales or larger, sometimes depositing its energy far outside of the host galaxy (e.g.~as in FR II radio sources).  GM11 showed that shock interaction between relativistic jet and the CNM powers non-thermal radio-infrared synchrotron radiation, which in the case of events viewed from a ``typical'' off-axis angle, peaks on a timescale $\sim$ months$-$years after disruption.  GM11 did not, however, consider the implications of events viewed face-on, as appears to characterize the geometry of \ubs~(Bloom et al.~2011) and \ubsb~\citep{Cenko+11}.

In this paper we extend the basic picture outlined by GM11 to develop an ``afterglow'' model for the low frequency (radio$-$microwave) emission from \ubs.  We begin in $\S\ref{sec:model}$ with a basic model for the jet-CNM interaction, using the observed prompt high energy emission to estimate the duration and power of the jet.  We argue that the radio emission from \ubs~is synchrotron radiation from behind the forward shock.  Using analytic expressions verified with numerical simulations, we show that the achromatic break observed in the radio light curve is naturally explained as the transition between an early phase when the reverse shock crosses the ejecta, to a later phase when the flow settles into a self-similar evolution.  Using the inferred synchrotron frequencies and the self-absorbed flux, in $\S\ref{sec:ubs}$ we constrain the parameters of the jet and the CNM.  One of our primary conclusions is that the radio emission alone provides robust constraints on the Lorentz factor and beaming fraction of the jet.  In $\S\ref{sec:predictions}$ we extend our model to make future predictions for the radio emission over the coming months and years.  Our one dimensional model is valid until the outflow decelerates to mildly relativistic speeds, at which point lateral expansion becomes important \citep{Zhang&Macfadyen09}.  We plan to explore this late transition to the Sedov phase, and the resulting predictions for the off-axis emission from events similar to \ubs, in greater detail in future work.  We summarize our conclusions in $\S\ref{sec:conclusions}$.  Although our analysis focuses on \ubs~since more data is available for this event, we conclude with a brief discussion of the implications of our model in the context of \ubsb.   

\section{Afterglow Model for TDE Jets}
\label{sec:model}

\begin{figure}
\resizebox{\hsize}{!}{\includegraphics[]{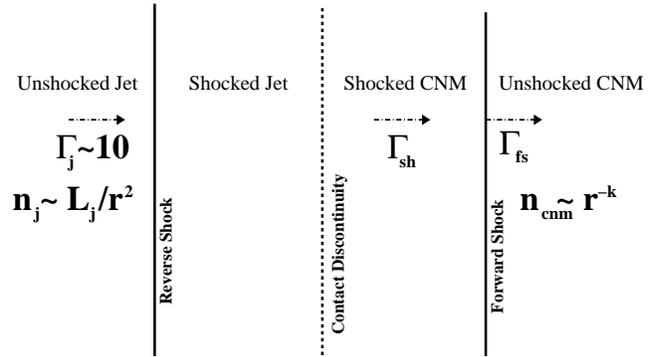}}
\caption[] {Schematic illustration of the forward-reverse shock structure produced by the interaction of a relativistic jet and the surrounding CNM during the early phase ($t \lesssim t_{\rm j}$) when the reverse shock crosses back through the shell of ejecta released during the initial period of jet activity.  The forward shock propagates with a Lorentz factor $\Gamma_{\rm fs}$ into the unshocked CNM (of density profile $n_{\rm cnm} \propto r^{-k}$), while a reverse shock propagates back into the unshocked jet (of density profile $n_{\rm j} \propto L_{\rm j}/r^{2}$; eq.~[\ref{eq:nj}]).  A contact discontinuity separates the shocked CNM from the shocked jet material.  Both the shocked jet and the shocked CNM move outwards with a similar Lorentz factor $\Gamma_{\rm sh}$, which may be calculated using the relativistic shock jump conditions assuming constant pressure across the contact (eq.~[\ref{eq:gshcross}]).} 
\label{fig:cartoon}
\end{figure}

\begin{figure}
\resizebox{\hsize}{!}{\includegraphics[]{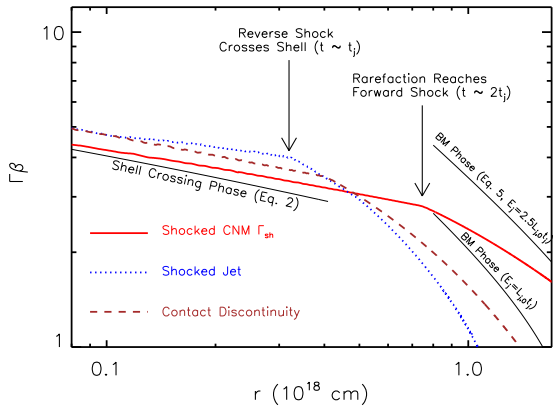}}
\caption[] {Lorentz factor of the shocked circumnuclear medium (CNM) behind the forward shock ($\Gamma_{\rm sh}$; {\it thick solid red line}), of the contact discontinuity ({\it dashed brown line}) and of the shocked jet ({\it dotted blue line}), extracted directly from a one-dimensional hydrodynamic simulation \citep{Mimica+09}.  The calculation is performed assuming a jet duration $t_{\rm j} = 3\times 10^{5}$ s, initial jet luminosity $L_{\rm j,0} = 10^{48}$ ergs s$^{-1}$ (see eq.~[\ref{eq:Lj}]) and CNM radial density profile $n_{\rm cnm} = 10(r/10^{18}{\,\rm cm})^{-1}$ cm$^{-3}$ ($k=1$).  Shown for comparison with black solid lines are the analytic approximations for $\Gamma_{\rm sh}$ during the early phase when the reverse shock is crossing through the shell of ejecta released during the initial period of constant jet luminosity ($t \lesssim t_{\rm j}$; eq.~[\ref{eq:gshcross}]), and at late times during the Blandford-McKee self-similar evolution ($t \gg 2t_{\rm j}$; eq.~[\ref{eq:gshBM}]) for different assumptions about the total energy of the blast wave $E_{\rm j,iso}$.  Note that a break occurs in the Lorentz factor of the shocked jet once the outflow reaches the radius $r = r_{\rm cross} \approx 3\times 10^{17}$ cm (observer time $t \approx t_{\rm j}$) at which the reverse shock has crossed entirely through the initial shell.  The break in $\Gamma_{\rm sh}$ from this transition occurs at a somewhat larger radius $r \approx 7\times 10^{17}$ cm ($t \approx 2t_{\rm j}$), once the rarefaction wave launched at $r = r_{\rm cross}$ reaches the forward shock.} 
\label{fig:compare}
\end{figure}

\begin{figure}
\resizebox{\hsize}{!}{\includegraphics[]{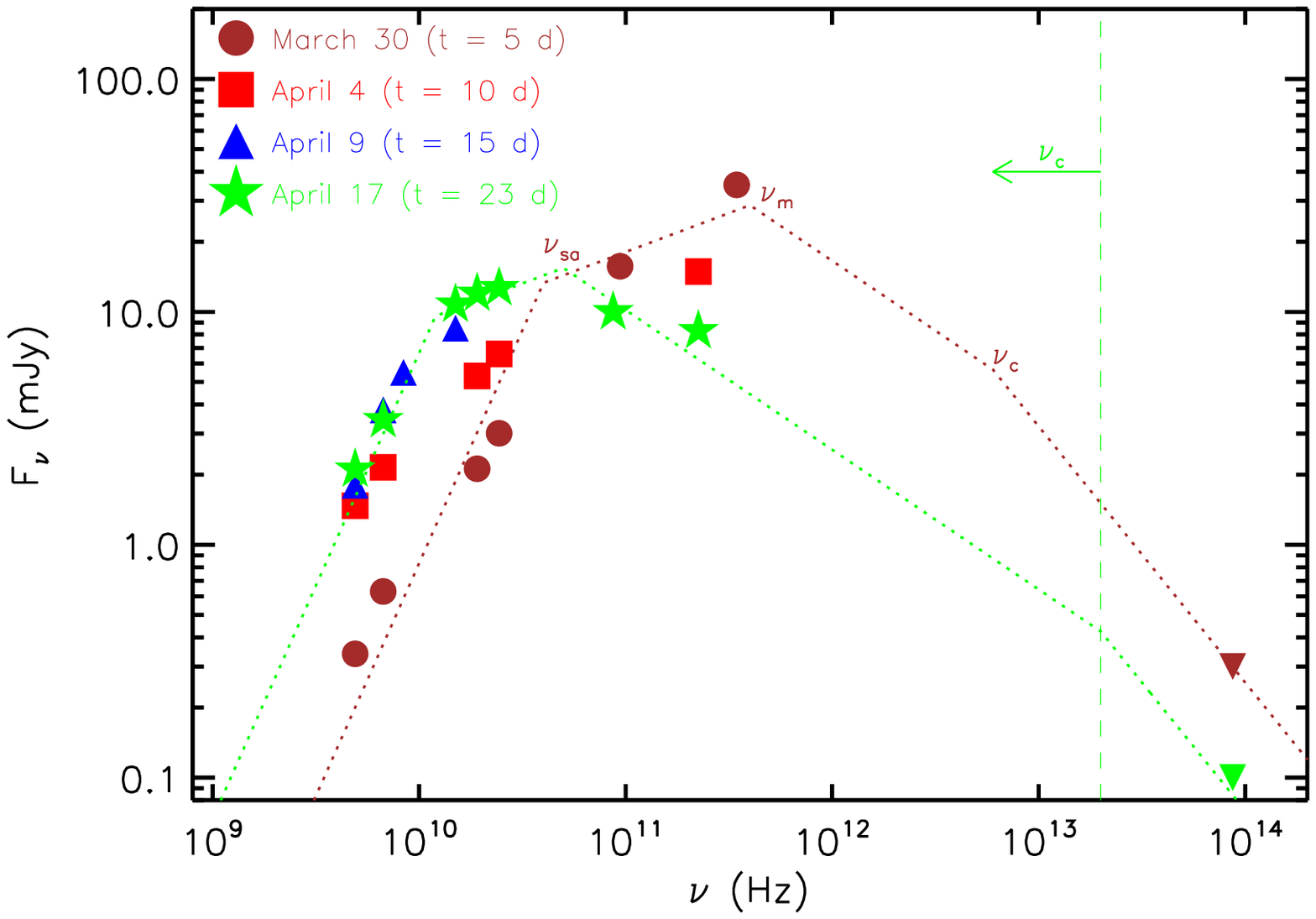}}
\caption[] {Radio$-$infrared spectra of \ubs, shown at various epochs with approximately contemporaneous data.  The radio data is from \citet{Zauderer+11}, while the NIR upper limits are L-band detections from \citet{Levan+11} (also Levan, private communication).  Overplot are model synchrotron spectra, fit to the data at the earliest and latest epochs, where we have assumed that the electron Lorentz factor distribution has a power-law index $p = 2.3$.  Notice that both the self-absorption frequency $\nu_{\rm sa}$ and the peak characteristic frequency $\nu_{\rm m}$ evolve to lower values with time, in a manner qualitatively consistent with the predictions in equations (\ref{eq:num}) and (\ref{eq:nusa}).} 
\label{fig:spectra}
\end{figure}

\begin{figure}
\resizebox{\hsize}{!}{\includegraphics[]{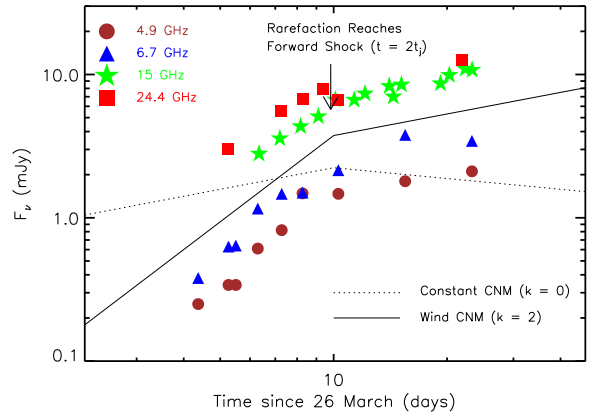}}
\caption[] {Radio light curves of \ubs~from \citet{Zauderer+11} at several frequencies, most of which are below the self-absorption frequency at most epochs (cf.~Fig.~\ref{fig:spectra}).  In our model the achromatic break in the light curve at $t \sim 10$ days occurs when the rarefaction wave (produced once the reverse shock crosses the initial shell of ejecta) catches up to the forward shock.  Shown for comparison are the predicted light curves (arbitrary normalization) if the CSM density has a constant ({\it dotted line}; $k = 0$) or a wind-type ({\it solid line}; $k = 2$) radial profile. } 
\label{fig:LC}
\end{figure}

We model the interaction between the jet and the CNM guided by previous work on GRB afterglows.  We adopt a simple model for the evolution of the (isotropic equivalent) kinetic power of the jet
\begin{equation}
L_{\rm j,iso}(\hat{t}) =
\left\{
\begin{array}{lr}
L_{\rm j,0}
, \qquad &
\hat{t} \le t_j \\
L_{\rm j,0}(\hat{t}/t_j)^{-5/3}
, \qquad &
\hat{t} > t_{\rm j}
\end{array}
\right. ,
\label{eq:Lj}
\end{equation}
motivated by the observed X-ray light curve of \ubs, which shows an initial period of several bright flares of similar luminosity, before a secular decline consistent with the expectation $L_{\rm j,iso} \propto \dot{M} \propto \hat{t}^{-5/3}$ \citep{Rees88} from fall-back accretion.\footnote{We place hats over the variable $\hat{t}$ to denote time in the central engine (lab) frame, thereby preserving the normal variable $t$ for ``observer time''.}  We adopt fiducial values for the initial luminosity $L_{\rm j,0} \equiv 10^{48}L_{48}$ erg s$^{-1}$ (see below) and jet duration\footnote{Note that the jet duration we adopt $t_{\rm j} \sim 3\times 10^{5}-10^{6}$ s is smaller than the fiducial value $t_j \sim 10^{7}$ s adopted by GM11, who assumed that the jet luminosity is constant at times when the fall-back rate is above the Eddington rate for a $\sim 10^{7}M_{\odot}$ BH.  GM11 also assumed that the pericenter radius $R_{\rm p}$ of the disrupted star was just interior to the tidal disruption radius $R_{t}$.  The shorter duration of the bright $\gamma$-/X-ray flaring from \ubs~instead suggests that either (1) the jet luminosity is directly proportional to the fall-back rate, regardless of the accretion Eddington factor, in which case $t_j$ instead represents the much shorter fall-back time; and/or (2) the stellar orbit was ``deeply plunging'' with a pericenter radius $R_{\rm p} \ll R_{\rm t}$ \citep{Cannizzo+11}.} $t_j \equiv 10^{6}t_{\rm j,6}$ s, as constrained from the total isotropic X/$\gamma$-ray energy $E_{\rm \gamma,iso} \equiv 3\times 10^{53}$ ergs measured over the first few weeks (e.g.~\citealt{Levan+11}; \citealt{Burrows+11}).  Although equation (\ref{eq:Lj}) obviously does not capture the full complexity of the observed emission, short-term irregularities in the jet power average out in the observed afterglow light curve (e.g.~\citealt{vanEerten+09}).

The inferred expansion rate of the radio emitting plasma from \ubs~suggests that the emission originates from a relativistic outflow (\citealt{Bloom+11}; \citealt{Zauderer+11}), while energetic arguments suggest that the emission was beamed into a bipolar jet.  Expanding on the latter point, if the mass of the disrupted star is $\sim 0.5M_{\odot}$ (near the peak of the stellar IMF or similar to that of a WD; \citealt{Krolik&Piran11}) and if $\lesssim$ half of its mass accretes following the disruption (\citealt{Rees88}; \citealt{Ayal+00}), then reconciling the total inferred energy of the jet $E_{\rm j,iso}^{\rm tot} \approx E_{\rm \gamma,iso}\epsilon_{\rm X}^{-1} \approx 6\times 10^{53}$ ergs with the accreted mass $M_{\rm acc} \sim M_{\odot}/4$ requires a beaming fraction $f_{\rm b} \equiv E_{\rm j}^{\rm tot}/E_{\rm j,iso}^{\rm tot}\sim 0.01(\epsilon_{\rm j}/0.01)$, where $E_{\rm j}^{\rm tot}$ is the true jet energy, $\epsilon_{\rm j} \equiv E_{\rm j}^{\rm tot}/M_{\rm acc}c^{2}$ is the fraction of the accreted rest mass used to power the jet, and $\epsilon_{\rm X} \sim 0.5$ is the radiative efficiency of the jet.  Motivated by statistics of blazar jets (e.g.~\citealt{Pushkarev+09}), theoretical constraints based on models of jet acceleration (e.g.~\citealt{Komissarov11}), and our detailed analysis of the radio data below ($\S\ref{sec:eta}$), we assume that the initial Lorentz factor (LF) of the jet obeys $\Gamma_{\rm j}\theta_{\rm j} \lesssim 1$, where $\theta_{\rm j}$ is the half opening-angle of the jet.  In this case the beaming fraction is set entirely by the $1/\Gamma_{\rm j}$ emission cone (independent of $\theta_{\rm j}$), such that the value of $f_{\rm b}$ required above corresponds to $\Gamma_{\rm j} \simeq (2f_{\rm b})^{-1/2} \sim 10(\epsilon_{\rm j}/0.01)^{-1/2}$.  In our more precise analysis in $\S\ref{sec:ubs}$, we determine the initial kinetic jet luminosity $L_{\rm j,0}$ (eq.~[\ref{eq:Lj}]) self-consistently by demanding that $E_{\rm j,iso} = (5/2)L_{\rm j,0}t_{\rm j} = 2E_{\rm \gamma,iso}(\Gamma_{\rm j}\theta_{\rm j})^{-2}$, where the factor $(\Gamma_{\rm j}\theta_{\rm j})^{2}/2 \lesssim 1$ corrects the observed isotropic emission to the isotropic kinetic energy of the jet.  This correction accounts for the fact that, although the jetted emission takes place within a cone of opening angle $\theta_{\rm obs} \sim 1/\Gamma_{\rm j}$, the jet energy is concentrated into the smaller physical jet opening angle ${\theta_{\rm j}}$ when $\theta_{\rm j}\Gamma_{\rm j} < 1$.    

We now describe the stages in the interaction between the jet and CNM using analytic expressions, which we will verify below with a full hydrodynamic simulation.  Initially after the jet forms, a forward shock (FS) is driven into the CNM, while a reverse shock (RS) simultaneously propagates back through the ejecta (see Fig.~\ref{fig:cartoon} for an illustration).  Jump conditions across the relativistic shocks and pressure balance at the contact discontinuity give an expression for the LF of the shocked fluid in the lab frame \citep{Sari&Piran95}:
\begin{eqnarray}
\Gamma_{\rm sh} = \Gamma_{\rm j}\left(1 + 2\Gamma_{\rm j}\left(\frac{n_{\rm cnm}}{n_{\rm j}}\right)^{1/2}\right)^{-1/2},
\label{eq:gshcross}
\end{eqnarray}
where $n_{\rm cnm}$ is the CNM density and
\be
n_{\rm j} = \frac{L_{\rm j, iso}}{4\pi r^{2}m_{p}c^{3}\Gamma_{\rm j}^{2}}
\label{eq:nj}
\ee
is the rest-frame density of the unshocked jet.  The LF of the FS is given by $\Gamma_{\rm fs} = \sqrt{2}\Gamma_{\rm sh}$, while the LF of the shocked material relative to the unshocked jet is given by
\be
\tilde{\Gamma}_{\rm rel} \simeq \frac{1}{2}\left(\frac{\Gamma_{\rm j}}{\Gamma_{\rm sh}} + \frac{\Gamma_{\rm sh}}{\Gamma_{\rm j}}\right),
\label{eq:grel}
\ee
where the tilde denotes a measurement in the frame of the unshocked jet.

Equations (\ref{eq:gshcross}) and (\ref{eq:grel}) show that the reverse shock is relativistic ($\tilde{\Gamma}_{\rm rel} \gg 1$) if $2\Gamma_{\rm j}(n_{\rm cnm}/n_{\rm j})^{-1/2} \gg 1$.  This condition,
\be
13 \left(\frac{L_{\rm j,iso}}{10^{48}\,\rm ergs\,s^{-1}}\right)^{-1/2}\left(\frac{n_{\rm cnm}}{\rm cm^{-3}}\right)^{1/2}\left(\frac{r}{10^{18}\,\rm cm}\right) \gg 1,
\label{eq:RSrelcondition}
\ee
is satisfied for reasonable values of the CNM density on radial scales probed by the jet during the first few weeks.  Below we assume that the RS is indeed at least mildly relativistic and show that this results in a self-consistent model for the radio emission from \ubs.

Assuming that the RS is (mildly) relativistic, after an observer time $t \sim t_j$ the RS crosses entirely through the shell of ejecta released during the initial period of constant jet luminosity ($\hat{t} \lesssim t_{\rm j}$).  After this point a rarefaction wave begins to propagate forward from the back of the shell (in effect ``communicating'' the end of the initial period of jet activity to the rest of the outflow).  After passing through the shocked ejecta and crossing the contact discontinuity, the rarefaction reaches the FS on a timescale $t \sim 2t_j$ (with the exact time depending weakly on the speed of the RS).  This results in a sudden change in the properties of the material behind the FS, which we will show below leads to a break in the synchrotron light curve.  Note that if the RS were not at least mildly relativistic, as we have assumed, then the observed break in the light curve break would in general occur on timescales much longer than $t_{\rm j}$.

Finally, at late times the outflow approaches the \citet{Blandford&McKee76} (hereafter BM76) self-similar evolution, for which the LF of the shocked fluid evolves as
\be
\Gamma_{\rm sh} = \left(\frac{17-4k}{16\pi}\frac{E_{\rm j,iso}}{m_{p}n_{\rm cnm}r^{3}c^{2}}\right)^{1/2}.
\label{eq:gshBM}
\ee
Note that this expression depends only on properties of the CNM and the total ejected energy $E_{\rm j,iso}$.  In the case of TDE, a factor of 1.5 times more energy is injected at times $\gtrsim t_{\rm j}$ by the $\propto \hat{t}^{-5/3}$ tail (eq.~[\ref{eq:Lj}]) than in the initial emission episode ($t \lesssim t_{\rm j}$).  Depending on when this late material in incorporated into the flow (which depends on its velocity with respect to that of the FS), $E_{\rm j,iso}$ will vary from $1$ to $2.5\times t_{\rm j}L_{\rm j,0}$.

Figure \ref{fig:compare} provides a comparison between our analytic expressions for $\Gamma_{\rm sh}(r)$ in equations \ref{eq:gshcross} and \ref{eq:gshBM} with the LF of the material behind the FS as determined directly from a one-dimensional hydrodynamic simulation of the jet-CNM interaction \citep{Mimica+09}.  In this calculation we have assumed a CNM density profile $n = 10(r/10^{18}\,{\rm cm})^{-1}$ cm$^{-3}$ (i.e.~k = 1) and use the jet power evolution $L_{\rm j,iso}(t)$ given in equation (\ref{eq:Lj}) for parameters $L_{\rm j,0} = 10^{48}$ ergs s$^{-1}$, $t_{\rm j} = 3\times 10^{5}$ s, and $\Gamma_{\rm j} = 10$.  

Figure \ref{fig:compare} shows that at early times ($t \lesssim t_{\rm j}$) the LF of the shocked CNM ({\it solid red line}) from the full simulation agrees well with the evolution $\Gamma_{\rm fs} \propto \Gamma_{\rm sh} \propto r^{-1/4}$ predicted by equation (\ref{eq:gshcross}) ({\it solid black line}) in both slope and normalization.  A clear break is observed in the LF of the shocked jet ({\it dotted blue line}) at the radius $r = r_{\rm cross} \approx 3\times 10^{17}$ cm at which the RS reaches the back of the initial shell of ejecta; as discussed above, this occurs at an observer time $t \approx 3\times 10^{5} $ s which is similar to the jet duration $t_{\rm j}$.  A rarefaction wave is also launched at this time, which reaches the FS at a somewhat larger radius $\sim 7\times 10^{17}$ cm (observer time $\simeq 2t_{\rm j}$), this time producing a break in the LF of the shocked CNM just behind the FS, which in turn produces a break in the FS light curve.   

Finally, at times $\gtrsim 2t_{\rm j}$, $\Gamma_{\rm sh}$ decreases somewhat more shallow with radius than the evolution $\Gamma_{\rm fs} \propto r^{-1}$ predicted by the BM76 self-similar (eq.~[\ref{eq:gshBM}]), due to the additional energy injection from the $\propto \hat{t}^{-5/3}$ tail (which in this example is assumed to have the same LF $\Gamma_{\rm j} = 10$ as the initial jet).  Note that the true late-time evolution of $\Gamma_{\rm sh}(r)$ lies between the BM76 evolution for a blast wave energy $2.5L_{\rm j,0}t_{\rm j}$ (with $\propto \hat{t}^{-5/3}$ tail) and $E_{\rm j,iso} = L_{\rm j,0}t_{\rm j}$ (without tail).  However, since below we are interested in comparing to data on timescales $\gtrsim 2 t_{\rm j}$, the no-tail BM76 solution ($E_{\rm j,iso} = L_{\rm j,0}t_{\rm j}$) is used since it provides a more accurate estimate of the value of $\Gamma_{\rm sh}$ on timescales relatively soon after the break.

\subsection{Modeling the emission from~\ubs}
\label{sec:ubs}

Figure \ref{fig:spectra} shows radio spectra of \ubs~at various epochs with approximately contemporaneous data from \citet{Zauderer+11}, with time measured starting from the first detection of X-ray activity on March 26 \citep{Krimm11}.  Note that at all epochs the spectrum is self-absorbed $F_{\nu} \propto \nu^{2}$ at the lowest frequencies $\nu \lesssim 10-100$ GHz.  This indicates that the source is relatively compact, consistent with its young age as inferred by the lack of evidence for previous activity from the location of \ubs.  Shown for comparison in Figure \ref{fig:spectra} are two model synchrotron spectra, fit to the earliest and latest epochs of radio data.  Note that both the characteristic (or peak) frequency $\nu_{\rm m}$ and the self-absorption frequency $\nu_{\rm sa}$ show evidence of decreasing with time.  Also shown are L-band detections in the near infrared (NIR) (with large error bars [not shown] due to possible dust extinction), which we treat as upper limits since it is unclear whether the NIR emission originates from the same source of synchrotron emission as what dominates at radio wavelengths.

Figure \ref{fig:LC} shows the radio light curves of \ubs~\citep{Zauderer+11}, focusing on several low frequency bands which are the least affected by scintillation.  The data are well fit by a power law $F_{\nu} \propto t^{\alpha}$ with $\alpha \approx 2$ at times $t \lesssim 10$ days, followed by break to a shallower rise $\alpha \approx 0.5$ at later times.  The apparently achromatic nature of this break suggests that it results from a change in the jet dynamics, rather than a spectral break passing through a particular bandpass (see also Fig.~\ref{fig:spectra}).  As we discuss in the next section, this evolution is consistent with the transition expected from the early shell-crossing phase to the late BM76 phase around the time $t \sim 2t_j \sim 10$ days, as illustrated in Figure \ref{fig:compare}.  Note that for \ubs~we adopt a value for the jet duration $t_{\rm j} \approx 5\times 10^{5}$ s ($t_{\rm j,6} = 0.5$).

\subsubsection{Evidence for a wind-type CNM}

Although electrons are (in principle) accelerated at both the FS and the RS, in $\S\ref{sec:eta}$ we argue that synchrotron emission from the FS likely dominates in the case of \ubs.  At frequencies below the self-absorption break ($\nu < \nu_{\rm sa}$), the radiated emission is given by the usual blackbody emissivity, calculated using a temperature $3kT \sim \gamma_m m_{e}c^{2}$ corresponding to that of the minimum LF $\gamma_{\rm m}$ of shock-accelerated electrons.  Here $\gamma_{\rm m} \simeq (p-2/p-1)(m_{p}/m_{e})\epsilon_{e}\Gamma_{\rm sh}$ follows from the shock jump conditions, $\epsilon_{e}$ is the fraction of the post-shock thermal energy in relativistic electrons, and $p \approx 2-3$ is the power-law index of the electron LF distribution.  The resulting source-frame specific luminosity is approximately given by 
\be
L_{\nu < \nu_{\rm sa}}(FS) = \frac{64\pi^{2}}{3}\frac{(p-2)}{(p-1)}\epsilon_{e}m_{p} r^{2}\nu^{2}\times (\Gamma_{\rm sh}\theta_{\rm j})^{2}/2,
\label{eq:Lnusa}
\ee
where the factor $(\Gamma_{\rm sh}\theta_{\rm j})^{2}/2 \le 1$ corrects for the missing flux due to the finite lateral extent of the jet. 

We now derive how the light curve $L_{\nu < \nu_{\rm sa}}$ scales with time both before and after the RS crosses the jet (ignoring the overall flux normalization for the moment), in order to constrain the density profile of the CNM.  If the CNM density scales with radius as a power law $n_{\rm cnm} \propto r^{-k}$ (Fig.~\ref{fig:cartoon}), then equations (\ref{eq:gshcross}) and (\ref{eq:gshBM}) show that $\Gamma_{\rm sh} \propto r^{(k-2)/4}[\propto r^{(k-3)/2}$] for the shell-crossing[Blandford-McKee] phases, respectively, where we have used equation (\ref{eq:nj}) and have assumed that $2\Gamma_{\rm j}\sqrt{n_{\rm cnm}/n_{\rm j}} \gg 1$ in equation (\ref{eq:gshcross}).  Observer time and radius are related by $t \simeq r/2c\Gamma_{\rm sh}^{2}$, such that $r \propto t^{2/(4-k)}[t^{1/(4-k)}]$.  From equation (\ref{eq:Lnusa}), we therefore find that $L_{\nu<\nu_{\rm sa}} \propto r^{2}\Gamma_{\rm sh}^{2} \propto r^{(k+2)/2}[r^{(k-1)}] \propto t^{(k+2)/(4-k)}[\propto t^{(k-1)/(4-k)}]$.  For $k = 0(1)[2]$ we see that the pre and post-break temporal indices are given by $\alpha_{1} = 0.5(1)[2]$ and $\alpha_{2} = -1/4(0)[1/2]$, respectively.  As the observed light curves in Figure \ref{fig:LC} clearly illustrate, the steep initial rise, and continued post-break rise, are both broadly consistent with the predictions for a wind-type CNM ($k \simeq 2$), but are clearly inconsistent with a constant CNM density ($k = 0$).\footnote{Note that we have again neglected the possibility of additional energy injection from the $\propto \hat{t}^{-5/3}$ tail of the jet luminosity, which may cause $\Gamma_{\rm sh}$ to decrease slower with radius (and hence the post-break light curve to rise faster with time) than is predicted by the constant-$E_{\rm j}$ BM76 evolution (see Fig.~\ref{fig:compare}).  However, given the observational uncertainty in the temporal slope $\alpha_{2}$ of the post-break light curve, and the theoretical uncertainty regarding when this energy will be incorporated into the forward shock (which depends on the LF of the jet at late times relative to that of the forward shock), this additional complication does not alter our basic conclusions.}     

\subsubsection{Wind CNM afterglow model}

Assuming that $k = 2$ with $n_{\rm cnm} = n_{18}(r/10^{18}\,\rm cm)^{-2}$ cm$^{-3}$, equations (\ref{eq:gshcross}) and (\ref{eq:gshBM}) can now be written
\begin{eqnarray}
\Gamma_{\rm sh} \approx
 \left\{
\begin{array}{lr}
4.2L_{48}^{1/4}n_{18}^{-1/4},
&
t \lesssim t_j \\
 10.9L_{48}^{1/2}t_{\rm j,6}^{1/2}n_{18}^{-1/2}r_{18}^{-1/2} \simeq 7.2 L_{48}^{1/4}n_{18}^{-1/4}\left(\frac{t}{t_{\rm j}}\right)^{-1/4},
&
t \gtrsim 2t_j \\
\end{array}
\right., 
\label{eq:gsh}
\end{eqnarray}  
where $n_{18}$ is in cm$^{-3}$ and in the last expression we again relate the emission radius $r \equiv r_{18}10^{18}$ cm and observer time by the expression $r \approx 2\Gamma_{\rm sh}^{2}ct/(1+z)$, viz.~
\begin{eqnarray}
r_{18} \approx
\left\{
\begin{array}{lr}
0.78\left(\frac{t}{t_{\rm j}}\right)t_{\rm j,6}L_{48}^{1/2}n_{18}^{-1/2},
&
t \lesssim t_j \\
2.3\left(\frac{t}{t_{\rm j}}\right)^{1/2}t_{\rm j,6}L_{48}^{1/2}n_{18}^{-1/2}
&
t \gtrsim 2t_j \\
\end{array}
\right., 
\label{eq:r18}
\end{eqnarray} 
Again note that we have assumed $E_{\rm j,iso} = L_{\rm j}t_{\rm j}$ (i.e.~neglecting the $\propto \hat{t}^{-5/3}$ tail) in the second expression in equation (\ref{eq:gsh}) since we are interested in the jet properties on timescales of $\lesssim$ few $\times t_{\rm j}$ (see Fig.~\ref{fig:compare}).

Based on the evolution for $\Gamma_{\rm sh}(t)$ in equation (\ref{eq:gsh}), we now give expressions for the time evolution of the characteristic synchrotron frequencies and the self-absorbed flux.  First, the peak (or characteristic) synchrotron frequency is given by 
\begin{eqnarray}
&\nu_{\rm m}& \simeq \frac{eB\gamma_{\rm m}^{2}\Gamma_{\rm sh}}{2\pi m_{e}c(1+z)} \nonumber \\
 &\approx & \left\{
\begin{array}{lr}
6\times 10^{10}{\,\rm Hz\,}\epsilon_{e,-1}^{2}\epsilon_{\rm B,-2}^{1/2}L_{48}^{1/2}t_{\rm j,6}^{-1}\left(\frac{t}{t_{\rm j}}\right)^{-1},
&
t \lesssim t_j \\
1.7\times 10^{11}{\,\rm Hz\,}\epsilon_{e,-1}^{2}\epsilon_{\rm B,-2}^{1/2}L_{48}^{1/2}t_{\rm j,6}^{-1}\left(\frac{t}{t_{\rm j}}\right)^{-3/2} ,
&
t \gtrsim 2t_j \\
\end{array}
\right., 
\label{eq:num}
\end{eqnarray}  
where $B = \sqrt{8\pi e_{\rm th}\epsilon_{\rm b}}$ and $e_{\rm th} \simeq 4\Gamma_{\rm sh}^{2}m_{p}n_{\rm cnm}c^{2}$ are the magnetic field and energy density of the shocked fluid, respectively; $\epsilon_{e} \equiv 0.1\epsilon_{e,-1}$; $\epsilon_{\rm B} \equiv 0.01\epsilon_{\rm B,-2}$; and we have assumed an electron LF power-law index $p = 2.3$.\footnote{Although we assume $p = 2.3$ throughout this paper, the value of $p$ always appears in the combination $\frac{(p-2)}{(p-1)}\epsilon_{e}$, such that one can adopt a different value of $p$ through a simple renormalization of the value of $\epsilon_{e}$.} 

Likewise, the time evolution of self-absorption frequency is given by (e.g.~\citealt{Granot&Sari02})
\begin{eqnarray} 
\nu_{\rm sa} \simeq 
\left\{
\begin{array}{lr}
3\times 10^{9}{\rm Hz\,}\epsilon_{e,-1}^{-1}\epsilon_{\rm B,-2}^{1/5}L_{48}^{-2/5}t_{\rm j,6}^{-1}n_{18}^{6/5}\left(\frac{t}{t_j}\right)^{-1}
&
t \lesssim t_j \\
6\times 10^{9}{\rm\,Hz\,}\epsilon_{e,-1}^{-1}\epsilon_{\rm B,-2}^{1/5}L_{48}^{-2/5}t_{\rm j,6}^{-1}n_{18}^{6/5}\left(\frac{t}{t_{\rm j}}\right)^{-3/5}
&
t \gtrsim 2t_j \\
\end{array}
\right., 
\label{eq:nusa}
\end{eqnarray} 
Equation (\ref{eq:nusa}) assumes that $\nu_{\rm sa} < \nu_{\rm m}$, which Figure \ref{fig:spectra} shows is valid for \ubs~at times $t \lesssim 23$ days (In particular the spectral slope above $\nu_{\rm sa} \approx 10-100$ GHz is consistent with $F_{\nu} \propto \nu^{1/3}$, but is much different than $F_{\nu} \propto \nu^{5/2}$, as would be expected if $\nu_{\rm m} < \nu_{\rm sa}$).  

Finally,  the flux of the self-absorbed emission at the distance of \ubs~($z \simeq 0.35; D_{L} = 5.7\times 10^{27}$ cm) is given by
\begin{eqnarray} 
&F_{\nu <\nu_{\rm sa}}& = \frac{L_{\nu}(\nu<\nu_{\rm sa})(1+z)^{3}}{4\pi D_{L}^{2}} \nonumber \\
&\approx& 
\left\{
\begin{array}{lr}
26{\,\rm mJy}\epsilon_{e,0.1}L_{48}^{3/2}t_{\rm j,6}^{2} n_{18}^{-3/2}\nu_{10}^{2}\theta_{j,-1}^{2}\left(\frac{t}{t_{\rm j}}\right)^{2}
&
t \lesssim t_j \\
240{\,\rm mJy}\epsilon_{e,0.1}L_{48}^{3/2}t_{\rm j,6}^{2}n_{18}^{-3/2}\nu_{10}^{2}\theta_{j,-1}^{2}\left(\frac{t}{t_{\rm j}}\right)^{1/2},
&
t \gtrsim 2t_j \\
\end{array}
\right., 
\label{eq:Fnusa}
\end{eqnarray}  
where $\theta_{\rm j} \equiv 0.1\theta_{j,-1}$, $\nu \equiv 10^{10}\nu_{10}$ Hz, and we have corrected the normalization of the flux in second expression to account for the detailed radial structure of the outflow, as described in the Appendix of \citet{Granot&Sari02}.


A casual comparison of the above expressions to the evolution of the model synchrotron spectra in Figure \ref{fig:spectra} shows good overall agreement.  For example, $\nu_{\rm m}$ is observed to decrease from $\gtrsim 3\times 10^{11}$ Hz at $t \simeq 5$ days $\sim t_{\rm j}$ to $\sim 3\times 10^{10}$ Hz at $t \simeq 23$ days $\approx 4t_{\rm j}$, consistent with the prediction $\nu_{\rm m} \propto t^{-3/2}$ for $t \gtrsim 2t_{\rm j}$ from equation (\ref{eq:num}).  Likewise, $\nu_{\rm sa}$ decreases from $\sim$ few$\times 10^{10}$ Hz to $\sim 10^{10}$ Hz, consistent with the slower evolution $\nu_{\rm sa} \propto t^{-3/5}$ predicted by equation (\ref{eq:nusa}).  

We now use the synchrotron frequencies and self-absorbed flux from the last epoch of data on April 17 ($t \approx 23$ d $\approx 4 t_{\rm j}$) to constrain the unknown parameters $\epsilon_{e}$, $n_{18}$, and $\Gamma_{\rm j}$.  As discussed near the beginning of this section, the observed high energy fluence constrains the isotropic jet luminosity to obey  
\begin{eqnarray} 
L_{48} \simeq 
\left\{
\begin{array}{lr}
0.5\eta^{-2},
&
\eta \le 1 \\
0.5,
&
\eta > 1 \\
\end{array}
\right., 
\label{eq:L48}
\end{eqnarray}  
where $\eta \equiv \Gamma_{\rm j}\theta_{\rm j}$.  Throughout this section we assume that $\eta \le 1$, although in $\S\ref{sec:eta}$ we consider (and argue against) the possibility that $\eta > 1$. 

\begin{figure}
\resizebox{\hsize}{!}{\includegraphics[]{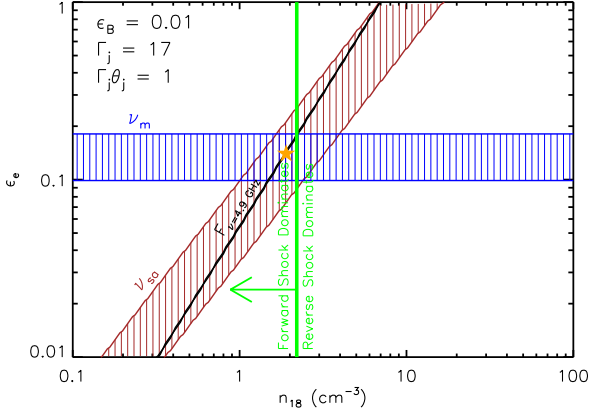}}
\caption[] {Constraints on the fraction of the shock energy placed into relativistic electrons ($\epsilon_{e}$) and the CNM density at $r = 10^{18}$ cm ($n_{18}$) from fitting the synchrotron spectrum from the final epoch of radio observations on April 17 ($t \approx 4t_{\rm j}$) to our FS model.   Constraints shown include the peak frequency $\nu_{\rm m}$ ({\it blue region}; eq.~[\ref{eq:num}]), the self-absorption frequency $\nu_{\rm sa}$ ({\it brown region}; eq.~[\ref{eq:nusa}]), and the self-absorbed flux at 4.9 GHz $F_{\nu = 4.9{\,\rm GHz}}$ ({\it black line}; eq.~[\ref{eq:Fnusa}]), including estimates of the uncertainties in the measured quantities.  In this example we take characteristic values for the fraction of the shock energy in magnetic fields $\epsilon_{\rm B} = 0.01$ and the parameter $\eta = \Gamma_{\rm j}\theta_{\rm j} = 1$, which results in a unique solution ({\it orange star}) for the values of $\epsilon_{e} \approx 0.15$, $n_{18} \approx 1.5$, and the initial LF of the jet $\Gamma_{\rm j} = 17$ (see also eqs.~[\ref{eq:eesolve}-\ref{eq:Gammajsolve}]).  The green line shows the upper limit on the density, such that the FS dominates over the RS emission as obtained using equation (\ref{eq:Lrel}).} 
\label{fig:constraint1}
\end{figure}

\begin{figure}
\resizebox{\hsize}{!}{\includegraphics[]{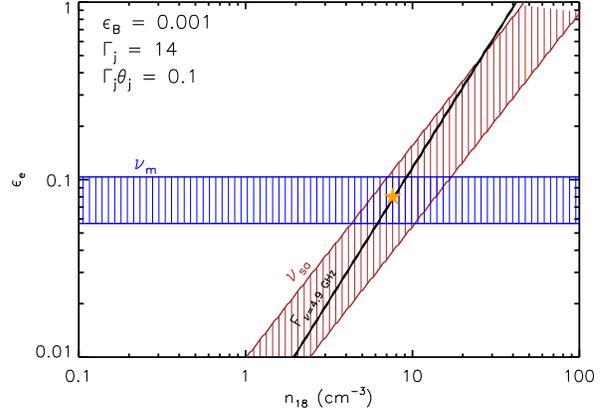}}
\caption[] {Same as Figure \ref{fig:constraint1}, but calculated for different values of $\epsilon_{\rm B} = 10^{-3}$ and $\eta = \Gamma_{\rm j}\theta_{\rm j} = 0.1$.} 
\label{fig:constraint2}
\end{figure}

Adopting fiducial values for the characteristic frequency $\nu_{\rm m}(t = 23$ d) $\equiv \nu_{\rm m,23} \approx 40$ GHz, self-absorption frequency $\nu_{\rm sa}(t = 23$ d) $\equiv \nu_{\rm sa,23} \approx$ 10 GHz and the 4.9 GHz flux $F_{\nu = 4.9{\,\rm GHz}}(t = 23$ d)$ \equiv F_{4.9,23} \approx$ 2 mJy motivated using the data in Figure \ref{fig:spectra}, equations (\ref{eq:num})$-$(\ref{eq:Fnusa}) can be solved for the following
\be
\epsilon_{e} = 0.12\epsilon_{\rm B,-2}^{-1/4}\left(\frac{\nu_{\rm m,23}}{10\,{\rm GHz}}\right)^{1/2}\eta^{1/2}
\label{eq:eesolve}
\ee
\be
n_{18} = 1.5\epsilon_{\rm B,-2}^{-3/8}\left(\frac{\nu_{\rm m,23}}{40\,{\rm GHz}}\right)^{5/12}\left(\frac{\nu_{\rm sa,23}}{10\,{\rm GHz}}\right)^{5/6}\eta^{-1/4}
\label{eq:n18solve}
\ee
\be
\Gamma_{\rm j} = 17\epsilon_{\rm B,-2}^{5/32}\left(\frac{\nu_{\rm m,23}}{40\,{\rm GHz}}\right)^{-1/16}\left(\frac{\nu_{\rm sa,23}}{10\,{\rm GHz}}\right)^{-5/8}\left(\frac{F_{4.9,23}}{2{\,\rm mJy}}\right)^{-1/2}\eta^{-1/16}.
\label{eq:Gammajsolve}
\ee

Equations (\ref{eq:eesolve})$-$(\ref{eq:Gammajsolve}) show that for reasonable values of $\epsilon_{\rm B} \approx 10^{-3}-0.1$ (e.g.~\citealt{Panaitescu&Kumar01}) and $\eta = \Gamma_{\rm j}\theta_{\rm j} \sim 0.1-1$ (e.g.~\citealt{Pushkarev+09}), one finds reasonable values for $\epsilon_{e} \approx 0.05-0.2$, $n_{18} \approx 0.3-10$, and $\Gamma_{\rm j} \simeq 10-20$.  Note that the value of $\Gamma_{\rm j}$ is particularly robust since it depends only weakly on the uncertain values of $\epsilon_{\rm B}$ and $\eta$.  As we discuss further in $\S$\ref{sec:conclusions}, this implies that the beaming fraction of \ubs~is well constrained to be $f_{\rm b} \sim 1/2\Gamma_{\rm j}^{2} \approx 1-5\times 10^{-3}$.

Figures \ref{fig:constraint1} and \ref{fig:constraint2} show the parameter space of $\epsilon_{e}$ and $n_{18}$ constrained by the observed values of $\nu_{\rm m,23}$, $\nu_{\rm sa,23}$, and $F_{4.9,23}$, including estimates of the uncertainties in these quantities.  We show two cases corresponding to different values of $\epsilon_{\rm B}$ and $\Gamma_{\rm j}$, the latter chosen from equation (\ref{eq:Gammajsolve}) to produce an acceptable solution for which the constraints on $\nu_{\rm sa}$, $\nu_{\rm m}$, and $F_{\nu = 4.9\rm GHz}$ all overlap.

\subsubsection{Constraints on the value of $\eta \equiv \Gamma_{\rm j}\theta_{\rm j}$}
\label{sec:eta}

Although $\Gamma_{\rm j}$ is well-constrained by equation (\ref{eq:Gammajsolve}), the combination $\eta = \Gamma_{\rm j}\theta_{\rm j}$ is poorly constrained by the analysis thus far.  In this section we present several independent arguments for why the value of $\eta$ is $\sim 1$.

We begin by placing a lower limit on $\eta.$  One constraint not yet considered is that for the RS to be at least mildly relativistic, as is necessary to explain the timing of the observed radio break, then the second term in parenthesis of Equation (\ref{eq:gshcross}) must be $\gg 1$ (see eq.~[\ref{eq:RSrelcondition}]).  Requiring that $\Gamma_{\rm sh} \ll \Gamma_{\rm j}$ at times $t \lesssim t_{\rm j}$ results in the following constraint:
\be
\eta = \theta_{\rm j}\Gamma_{\rm j} > 0.05\epsilon_{\rm B,-2}^{-1/6}\left(\frac{\nu_{\rm m,23}}{40\,{\rm GHz}}\right)^{-17/54}\left(\frac{\nu_{\rm sa,23}}{10\,{\rm GHz}}\right)^{10/9}\left(\frac{F_{4.9,23}}{2{\,\rm mJy}}\right)^{4/3},
\label{eq:etamin}
\ee
where we have used equations (\ref{eq:gsh}), (\ref{eq:n18solve}) and (\ref{eq:Gammajsolve}).

Although the constraint in equation (\ref{eq:etamin}) is relatively weak, one can motivate a somewhat higher value for $\eta$ by considering the physical value of CNM density encountered by the shock at the last epoch of observations.  Using equation (\ref{eq:r18}) we find that
\be
n_{\rm cnm} \simeq 0.4n_{18}^{2}\eta^{2}\left(\frac{t}{4t_{\rm j}}\right)^{-1} {\rm\,cm^{-3}},
\label{eq:ncnmphysical}
\ee
which shows that if $\eta \sim 0.1$ (implying $n_{18} \lesssim 6$ for $\epsilon_{\rm B} > 10^{-3}$ using eq.~[\ref{eq:n18solve}]), then the density probed at $t \approx 4t_{\rm j}$ is $n_{\rm cnm} \lesssim 0.1$ cm$^{-3}$.  If this value is accurate, then the CNM density in the host of \ubs~ is much lower than the density $\gtrsim 10$ cm$^{-3}$ inferred on a similar radial scale in the center of the Milky Way (\citealt{Baganoff+03}; \citealt{Quataert04}).  Such a low density would be particularly surprising because several magnitudes of dust extinction were inferred along the line of sight to \ubs~\citep{Bloom+11}.  We conclude that the unphysically low CNM density implied by $\eta \ll 1$ instead favors a value for $\eta \gtrsim 0.3$.  

An upper limit on the value of $\eta$ can be placed by requiring that emission from behind the FS dominates that from the RS.  The light curves of both the FS and the RS at frequencies $\nu < \nu_{\rm sa}$ initially rise as $F_{\nu} \propto t^{2}$ for a wind-type CNM, such that the data at times $t \lesssim t_{\rm j}$ cannot alone be used to determine which dominates the emission.  After the shock crosses the ejecta ($t \gtrsim t_{\rm j}$), however, the RS emission is predicted to decrease more rapidly (or increase less rapidly) than is observed in \ubs~.  This is because no new electrons are accelerated at the RS at times $t \gtrsim t_{\rm j}$, yet the plasma loses thermal energy to adiabatic expansion.  In particular, emission from the RS at $\nu \gtrsim \nu_{\rm m}$ is predicted to decrease as $F_{\nu} \propto t^{-2}$ or faster (e.g.~\citealt{Sari&Piran99}), contrary to the observed high frequency post-break light curve of \ubs~\citep{Zauderer+11}.  

On timescales $t \sim t_{\rm j}$ the brightness of the forward shock relative to that of the RS is given by the ratio of the relative electron temperatures (see eq.~[\ref{eq:Lnusa}]), which is in turn proportional to the ratio of the relative LFs 
\be
\frac{L_{\nu < \nu_{\rm sa}}(FS)}{L_{\nu < \nu_{\rm sa}}(RS)} \simeq \frac{\Gamma_{\rm sh}}{\tilde{\Gamma}_{\rm rel}} \approx \frac{2\Gamma_{\rm sh}^{2}}{\Gamma_{\rm j}} \approx 2.5 n_{18}^{-1/2}\left(\frac{\Gamma_{\rm j}}{10}\right)^{-1},
\label{eq:Lrel}
\ee
where we have used equations (\ref{eq:grel}) and (\ref{eq:gsh}) for $L_{48} = 0.5$ (eq.~[\ref{eq:L48}]).  Again solving for the unknowns as in equations (\ref{eq:eesolve}-\ref{eq:Gammajsolve}) above, but this time assuming $\eta = \Gamma_{\rm j}\theta_{\rm j} > 1$, we now find
\be
\epsilon_{e} = 0.12\epsilon_{\rm B,-2}^{-1/4}\left(\frac{\nu_{\rm m,23}}{30\,{\rm GHz}}\right)^{1/2}
\label{eq:eesolve2}
\ee
\be
n_{18} = 1.5\epsilon_{\rm B,-2}^{-3/8}\left(\frac{\nu_{\rm m,23}}{40\,{\rm GHz}}\right)^{5/12}\left(\frac{\nu_{\rm sa,23}}{10\,{\rm GHz}}\right)^{5/6}
\label{eq:n18solve2}
\ee
\be
\theta_{\rm j} = 0.05\epsilon_{\rm B,-2}^{-5/32}\left(\frac{\nu_{\rm m,23}}{40\,{\rm GHz}}\right)^{1/16}\left(\frac{\nu_{\rm sa,23}}{10\,{\rm GHz}}\right)^{5/8}\left(\frac{F_{4.9,23}}{2{\,\rm mJy}}\right)^{1/2},
\label{eq:thetajsolve}
\ee
Substituting the expressions for $n_{18}$ and $\theta_{\rm j}$ into equation (\ref{eq:Lrel}) gives 
\begin{eqnarray}
& &\frac{L_{\nu < \nu_{\rm sa}}(FS)}{L_{\nu < \nu_{\rm sa}}(RS)} \simeq \nonumber \\
& &1.3\epsilon_{\rm B,-2}^{1/32}\left(\frac{\nu_{\rm m,23}}{40\,{\rm GHz}}\right)^{-7/48}\left(\frac{\nu_{\rm sa,23}}{10\,{\rm GHz}}\right)^{5/24}\left(\frac{F_{4.9,23}}{2{\,\rm mJy}}\right)^{1/2}(\Gamma_{\rm j}\theta_{\rm j})^{-1}.
\label{eq:Lrel2}
\end{eqnarray}
For the FS to dominate over the RS (as observed) we thus require $\eta = \Gamma_{\rm j}\theta_{\rm j} \lesssim 1$.    

To summarize, the available data constrain $\eta =  \Gamma_{\rm j}\theta_{\rm j}$ to lie the range $\sim 0.05-1$, but given a realistic lower limit on the CNM density $\eta$ most likely lies in the range $\sim 0.3-1$.  These values of $\eta$ are consistent with those inferred from observations of blazar jets (e.g.~\citealt{Pushkarev+09}) and those expected from theoretical models of relativistic jet acceleration (e.g.~\citealt{Komissarov11}).

\section{Future Evolution}
\label{sec:predictions}
\subsection{Ongoing Blandford-McKee Phase}
\label{sec:BMfuture}
\begin{figure}
\resizebox{\hsize}{!}{\includegraphics[]{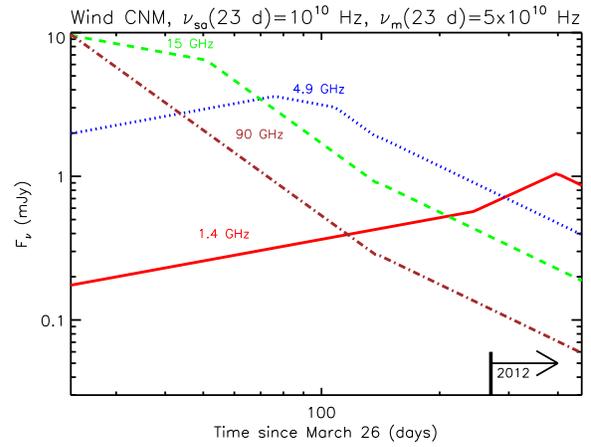}}
\caption[] {Future evolution of the radio light curves of \ubs, assuming that $\nu_{\rm sa} = 10^{10}$ Hz and $\nu_{\rm m} = 5\times 10^{10}$ Hz at $t = 23$ days, and that the CNM continues to maintain a wind-type radial density profile $n_{\rm cnm} \propto r^{-2}$.  } 
\label{fig:future1}
\end{figure}

\begin{figure}
\resizebox{\hsize}{!}{\includegraphics[]{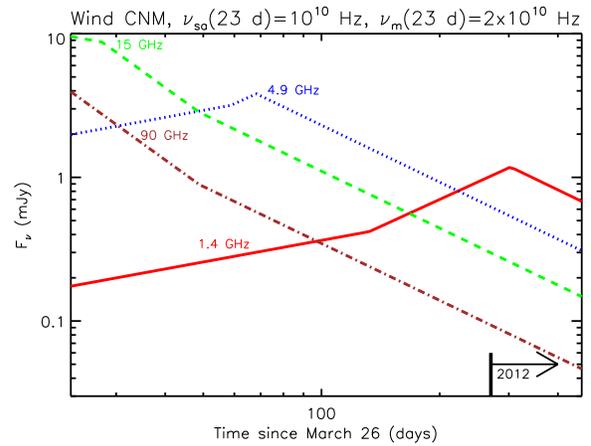}}
\caption[] {Same as Figure \ref{fig:future1}, but assuming that $\nu_{\rm m} = 2\times 10^{10}$ Hz at $t = 23$ days.  Taking into account uncertainty induced by scintillation at high frequencies, note that the flux at 90 GHz does not match precisely onto its observed value at $t = 23$ days in this case.} 
\label{fig:future2}
\end{figure}

\begin{figure}
\resizebox{\hsize}{!}{\includegraphics[]{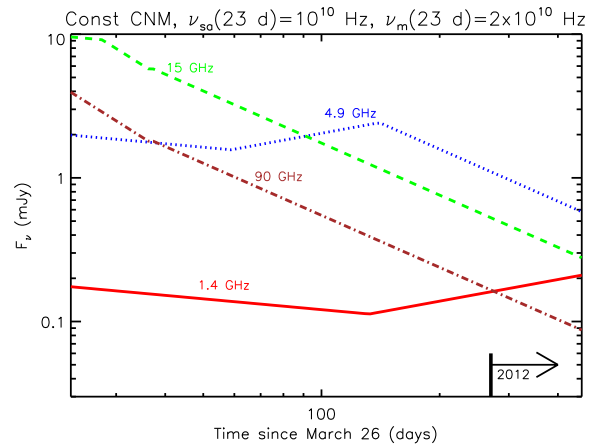}}
\caption[] {Same as Figure \ref{fig:future2}, but instead assuming that the CNM density profile flattens to a constant value at times $t \gtrsim 23$ days.} 
\label{fig:future3}
\end{figure}

In this section we extend our model for the synchrotron emission from \ubs~to predict the ongoing evolution of the radio light curves over the coming months to years.  If the outflow continues to evolve as predicted by the BM76 evolution with constant energy in a wind-type CNM (eq.~[\ref{eq:gshBM}]), then the self-absorbed flux and peak frequency will continue to evolve as $F_{\nu < \nu_{\rm sa}} \propto t^{1/2}$ (eq.~[\ref{eq:Fnusa}]) and $\nu_{\rm m} \propto t^{-3/2}$ (eq.~[\ref{eq:num}]), respectively. The self absorption frequency will continue to evolve as $\nu_{\rm sa} \propto t^{3/5}$ while $\nu_{\rm sa} < \nu_{\rm m}$, but then will change its temporal slope slightly to $\nu_{\rm sa} \propto t^{-(3p+5)/(2p+10)}$ once $\nu_{\rm sa} > \nu_{\rm m}$ \citep{Granot&Sari02}.  

Using this information alone, Figures \ref{fig:future1} and \ref{fig:future2} shows predictions for the future evolution of the radio flux at several common observing frequencies, for two different assumptions about the peak synchrotron frequency at $t = 23$ days (which, as shown in Fig.~\ref{fig:spectra}, is somewhat uncertain).  At high frequencies $\gtrsim 10$ GHz, the flux will continue to decrease monotonically with time, approaching $\lesssim 0.1$ mJy at 90 GHz on timescales of a year.  At low frequencies the evolution is more interesting: at 1.4 and 4.9 GHz the flux should initially continue to rise with $F_{\nu} \propto t^{1/2}$.  Then, on a timescale of months, the peak synchrotron frequency will decrease below the self-absorption frequency, after which point the spectrum is rises as $F_{\nu} \propto \nu^{5/2}$ for $\nu_{\rm m} < \nu < \nu_{\rm sa}$.  When an observed bandpass falls between this frequency range, the flux is predicted to rise as $F_{\nu} \propto t^{5/4}$.  Since this condition is likely to be satisfied at 1.4 GHz (and possibly at 4.9 GHz), the temporal slope of the light curve in these wavebands may suddenly increase by a value $\gtrsim 1/2$ after a timescale of months.

Although the CNM density inferred from the early radio data is consistent with a wind-type radial profile, the CNM profile could in principle change with radius, in particular considering that very low CNM densities are implied by the wind-like profile at late times (see eq.~[\ref{eq:ncnmphysical}] and surrounding discussion).  In order to explore this possibility, we also consider how the radio evolution would change if the CNM density flattens to a constant value at times $\gtrsim 4 t_{\rm j} \approx $ 23 days.  In this case, the evolution can be calculated in the same manner as above, but using scalings for the synchrotron flux and frequencies which are instead appropriate for a constant CNM ($k = 0$): $F_{\nu < \nu_{\rm sa}} \propto t^{-1/4}$, $\nu_{\rm m} \propto t^{-3/2}$, $\nu_{\rm sa} \propto t^{0}$ (when $\nu_{\rm sa} < \nu_{\rm m}$), $\nu_{\rm sa} \propto t^{-(3p+1)/(2p+10)}$ (when $\nu_{\rm sa} > \nu_{\rm m}$).  Figure \ref{fig:future3} shows our results in the constant CNM case, which show that initially the flux at all frequencies begins to decrease.  However, this time when the self-absorption frequency crosses below the characteristic frequency, a low frequency {\it rebrightening} occurs on a timescale of months.   

Thus far we have neglected the effects of the $\propto \hat{t}^{-5/3}$ tail of the jet luminosity (eq.~[\ref{eq:Lj}]) on the late-time radio emission.  Depending on when this material catches up to the forward shock, energy deposition from the tail may cause $\Gamma_{\rm sh}$ to decrease less rapidly with time than would be predicted with no additional energy injection (see Fig.~\ref{fig:compare}).  The effect of the tail in this case may be to increase the temporal indices of the light curves above those predicted in Figures \ref{fig:future1}-\ref{fig:future3}.  Though more speculative, another possible source of late emission may occur from the RS, if it continue to propagate back through the ejecta in the tail after passing through the initial ejecta shell (see \citealt{Uhm&Beloborodov07} and \citealt{Genet+07} for such a model applied to normal GRB afterglows).  A detailed exploration of this idea is, however, beyond the scope of this work.

\subsection{Sedov Phase}
\label{sec:STfuture}

Our calculation thus far assumed an ultra-relativistic outflow, which allowed us the simplification of treating the flow as essentially one-dimensional.  Recent 2D hydrodynamical simulations of deceleration of a collimated blastwave demonstrate that the flow remains approximately conical until the LF of the FS decreases to $\lesssim 2$ (\citealt{Zhang&Macfadyen09}; cf.~\citealt{Wygoda+11}) at a time $t \approx t_{\rm NR}$.  This conclusion is supported by recent analytic models (\citealt{Lyutikov11}; \citealt{Granot&Piran11}).  After a transient phase of lateral expansion on timescales $t \gtrsim t_{\rm NR}$, the blast will ultimately relax into a non-relativistic, spherical expansion (Sedov-Taylor phase) centered at the location of the deceleration of the blast (GM11).

Using equation (\ref{eq:gsh}) (now increased by a factor 2.5$^{1/2} \approx 1.6$ to now account for the additional energy injected by the $\propto \hat{t}^{-5/3}$ tail) we estimate that the non-relativistic transition (taken to be $\Gamma_{\rm sh} \lesssim 2$) will occur at a time
\be
t_{\rm NR}(k=2) \simeq 8.8{\,\rm yr\,}\eta^{-2}n_{18}^{-1} \approx 5.9{\,\rm yr\,}\epsilon_{\rm B,-2}^{3/8}\eta^{-7/4},
\label{eq:tNR}
\ee
where in the second expression we have substitute equation (\ref{eq:n18solve}) for $n_{18}$.  Depending on the value of $\eta$, the Sedov phase is thus predicted to begin on a timescales of years to a decades.  

The non-relativistic transition may actually occur somewhat earlier than predicted by equation (\ref{eq:tNR}) if the density of the CNM does not decrease $\propto r^{-2}$ all the way out to the Sedov radius, as would be expected if the ambient ISM places a density ``floor'' $n_{\rm cnm} \gtrsim n_{\rm ism} \sim 0.1-1$ cm$^{-3}$.  Substituting $t_{\rm NR}$ into equation (\ref{eq:ncnmphysical}), we infer that the physical density at the Sedov radius is $n_{\rm cnm}(t_{\rm NR}) \sim 3\times 10^{-3}n_{18}^{3}$ cm$^{-3}$ for $\eta \sim 1$.  Thus, depending on the values of $n_{18}$ and $n_{\rm ism}$, the non-relativistic transition may instead at the usual Sedov time for a constant ambient density:
\be
t_{\rm NR}(k=0) \simeq 3.6{\rm yr\,}\eta^{-2/3}\left(\frac{n_{\rm ism}}{\rm cm^{-3}}\right)^{-1/3},
\ee   
somewhat earlier than predicted by equation (\ref{eq:tNR})

The angular size of the radio emission at the non-relativistic transition can be estimated to be
\be
\Theta \equiv \frac{2\theta_{\rm j}r(t = t_{\rm NR})}{D_{\rm A}} \approx 0.16 \theta_{j,-1}n_{18}^{-1} {\rm\,mas}, 
\label{eq:resolve}
\ee
where $D_{\rm A} \approx 1.0$ Gpc is the angular diameter distance to \ubs~.  Here we have used equation (\ref{eq:r18}) to estimate the jet radius at the non-relativistic transition $r(t = t_{\rm NR})$, assuming (optimistically) that the latter occurs at the time given by equation (\ref{eq:tNR}).  Unfortunately, even for optimistic assumptions, the angular scale implied by (\ref{eq:resolve}) is unlikely to be resolvable with VLBI, especially given the low predicted flux at high frequencies at late times (Fig.~\ref{fig:future1}-\ref{fig:future3}).

\section{Conclusions}
\label{sec:conclusions}

Building on the work of GM11, in this paper we have developed a model for the radio emission from \ubs~produced by the shock interaction between the jet and the gaseous CNM.  We have shown that the achromatic break in the observed light curve (Fig.~\ref{fig:LC}) is well explained as the transition from the early reverse shock crossing phase to the late BM76 evolution (Fig.~\ref{fig:compare}) if the CNM has a wind-like radial profile (Fig.~\ref{fig:LC}).  One implication of a wind-like CNM is that the post-shock LF is constant during the shell-crossing phase ($t \lesssim t_{\rm j}$) and only decreases slowly with time $\Gamma_{\rm sh} \propto t^{-1/4}$ thereafter (eq.~[\ref{eq:gsh}]).  The approximately fixed value of $\Gamma_{\rm sh} \sim$ few that we find over the first few weeks is one explanation for the approximately constant expansion velocity inferred by \citet{Zauderer+11}.  We emphasize, however, that the initial LF of the unshocked jet $\Gamma_{\rm j}$ is much higher than its post-shock value.  Indeed, a value of $\Gamma_{\rm j} \gg 1$ is required for beaming of the initial X-ray emission, as is necessary to reconcile the isotropic energy of \ubs~with the accretion of a stellar mass object. 

Achromatic breaks in afterglow light curves can in principle also occur when the LF of the shocked jet $\Gamma_{\rm sh}$ decreases below $\sim 1/\theta_{\rm j}$ (a so-called ``jet break''), as is commonly discussed in the context of GRBs (e.g.~\citealt{Frail+01}).  A standard jet break cannot, however, explain the radio break observed in \ubs~for three reasons.  First, as discussed in $\S\ref{sec:eta}$, the absence of dominant emission from the RS constrains the value of $\eta = \Gamma_{\rm j}\theta_{\rm j}$ to be $\lesssim 1$ (eq.~[\ref{eq:Lrel2}]), in which case $\Gamma_{\rm sh}\theta_{\rm j}$ is $\lesssim 1$ even at early times.  Second, for a wind-type CNM (as inferred from the pre-break light curve), no change in the temporal slope is predicted after a putative jet break because $\Gamma_{\rm sh}$ is constant with time during the RS crossing phase.  A jet break that occurs after the RS crossing is ruled out because it would then be the {\it second} break and only one break is observed.  Finally, if the observed radio break were in fact a jet break, then its occurrence at the time $t \approx 10$ d $\approx 2t_{\rm j}$ would be entirely coincidental.  In our model the observed time of the break is naturally explained as when the rarefaction wave reaches the forward shock (Fig.~\ref{fig:compare}). 

Using the inferred synchrotron frequencies and the self-absorbed flux, our analysis results in reasonable values for the fraction of the post-shock thermal energy in relativistic electrons $\epsilon_{e} \approx 0.05-0.2$; the CNM density at $10^{18}$ cm $n_{\rm cnm} \approx 1-10$ cm$^{-3}$; and the initial LF of the jet $\Gamma_{\rm j} \sim 10-20$ (eq.~[\ref{eq:eesolve}]-[\ref{eq:Gammajsolve}]).  Although the jet opening angle is less well-constrained, it most likely lies in the range $\theta_{\rm j} \sim (0.3-1)\times \Gamma_{\rm j}^{-1} \sim 0.01-0.1$ ($\S\ref{sec:eta}$), with much lower values disfavored since this requires an unphysically low value of the CNM density (eq.~[\ref{eq:ncnmphysical}]).  Although our model makes several simplifying assumptions (a jet with a sharp edge; broken power-law spectra; constant microphysical parameters), it is nevertheless gratifying that the properties of the relativistic outflow that we infer are remarkably consistent with those thought to characterize normal blazar and AGN jets. 

Perhaps our most important conclusion is that the radio emission from \ubs~provides evidence for a narrowly collimated jet, which is independent of (and stronger than) other arguments based on rates or total energetics.  In particular, the beaming fraction of the jet $f_{\rm b} = \theta_{\rm b}^{2}/2$ where $\theta_{\rm b} = {\rm max}[\Gamma_{\rm j}^{-1},\theta_{\rm j}]$ is constrained to lie in the relatively narrow range $\sim 1-5\times 10^{-3}$ (see eqs.~[\ref{eq:Gammajsolve}] and [\ref{eq:thetajsolve}]).  This conclusion is robust and independent of the assumed value for $\eta = \Gamma_{\rm j}\theta_{\rm j}$.  

The beaming fraction that we infer for the jetted emission from \ubs~has several implications.  First, it implies that the true beaming-corrected peak luminosity of the prompt X-ray/$\gamma-$ray emission may be as low as $\sim 10^{45}$ erg s$^{-1}$, similar to the Eddington luminosity of a $10^{7}M_{\odot}$ SMBH.  Super-Eddington accretion may thus not be required to explain the observed X-ray emission, whether it originates internal to the jet or from the disk itself (\citealt{Socrates11}).  On the other hand, the beaming-corrected energy of the event was $\sim 10^{51}$ ergs, which in the case of jetted emission requires a jet efficiency of $\epsilon_{\rm j} \approx 10^{-2}$ if the event was indeed powered by the accretion of a solar-mass star.  Note also that the beaming fraction we infer appears to be larger than that required to reconcile the rate of \ubs-like events with the TDE rate inferred by independent means (e.g.~\citealt{Wang&Merritt04}; \citealt{vanVelzen+11}; \citealt{Bower11}; \citealt{Kesden11}) by a factor of $\sim 10-30$ (see discussions in \citealt{Burrows+11}, \citealt{Cenko+11}).  This suggests that only a small fraction of TDE events are accompanied by [detectable] relativistic outflows.  Finally, the narrow beaming angle that we infer may in principle also constrain the spin of the SMBH via the [lack of] spin-induced precession inferred by the approximately steady jet luminosity over the first several months of observations (\citealt{Stone&Loeb11}; cf.~\citealt{Lei&Zhang11}).  

In $\S\ref{sec:predictions}$ we make predictions for the future evolution of the radio$-$microwave light curve of \ubs~over the coming months$-$years (Fig.~\ref{fig:future1}-\ref{fig:future3}).  At high frequencies the radio emission will continue to decrease with time, while at low frequencies the predicted light curves are initially relatively flat, depending on whether the CNM density continues to be well-described by a $\propto r^{-2}$ radial profile (or whether the profile flattens) and when/whether additional energy reaches the FS from the $\propto \hat{t}^{-5/3}$ tail.  Interestingly, at the lowest frequencies $\lesssim$ few GHz we predict that the flux may begin to increase more rapidly (or even rebrighten) on a timescale of months, once the characteristic frequency decreases below the self-absorption frequency (indeed, such a transition may already have begun).  These predictions can be used to distinguish our afterglow model for the radio emission powered by external CNM interaction, from other models which invoke synchrotron emission entirely internal to the jet itself (\citealt{vanVelzen+11}; \citealt{Miller&Gultekin11}).

An important theme of this work is that relativistic outflows from tidal disruption events provide a unique probe of the conditions in distant, {\it inactive} galactic nuclei.  Such observations are rare, except in the case Sgr A$^{\star}$ and nearby low-luminosity AGN, and are unprecedented in galaxies at high redshift.  If the wind-type CNM that we infer from \ubs~truly indicates an outflow from the nucleus of the host galaxy, then the mass loss rate [prior to the TDE] may be estimated as
\be
\dot{M} = 4\pi r^{2}v_{\rm w}n_{\rm cnm}m_{p} \simeq 3\times 10^{-5}M_{\odot}\,{\rm yr^{-1}}n_{18}\left(\frac{v_{\rm w}}{300\,\rm km\,s^{-1}}\right),
\label{eq:mdot}
\ee
where $v_{\rm w}$ is the wind velocity, scaled to a value characteristic of the expected escape speed from the SMBH at radii $\sim 10^{18}$ cm probed by the radio afterglow.  Depending on the value of $n_{18} \sim 1-10$, this mass loss rate is similar to that expected from $\sim 1-10$ Wolf-Rayet (WR) stars.  If scaled to the $\lesssim 10^{3}$ total number of WR stars in the Milky Way, this would suggest that the steady-state star formation rate is $\dot{M}_{\star} \sim 10^{-3}-10^{-2} M_{\odot}$ yr$^{-1}$ within the inner pc (given the Milky Way SFR rate $\sim M_{\odot}$ yr$^{-1}$).  Alternatively, our inference that the mass loss rate is similar to that of a massive star is also consistent with the model of \citet{Quataert&Kasen11}, who argue that \ubs~may simply be an unusual long-duration GRB associated with a stellar core collapse event.  

Finally, given the uncertainties, the radio data may also be consistent with an $n_{\rm cnm} \propto r^{-3/2}$ density profile, as predicted for Bondi accretion.  In this case the {\it accretion} rate that one infers is similar to the outflow rate given in equation (\ref{eq:mdot}).  Interpreted this way, $\dot{M}$ is several orders of magnitude lower than the accretion rate required to explain the outburst from \ubs.  This would provide additional evidence that \ubs~resulted from a very large increase in the SMBH accretion rate over its long-term average, as likely can only be explained by the tidal disruption of a star.
  
Although we have focused on modeling the emission from \ubs, our results can also be applied to the other recent jetted TDE candidate \ubsb~\citep{Cenko+11}.  The X-ray and radio brightness of \ubsb~is similar to that of \ubs~at a similar epoch, except that no spectral steepening is observed at low frequencies, suggesting that the self-absorption frequency obeys $\nu_{\rm sa} \lesssim 3\times 10^{9}$ Hz.  Equation (\ref{eq:n18solve}) shows that one way this observation could be explained in our model is if the CNM density is lower by a factor of a few than in the case of \ubs, possibly consistent with the lower optical extinction inferred towards \ubsb.

\section*{Acknowledgments}
We thank A.~Levan, J.~Bloom, B.~Cenko, A.~Socrates, and N.~Stone for helpful conversations.  We thank the anonymous reviewer for helpful comments and suggestions.  BDM is supported by NASA through Einstein Postdoctoral Fellowship grant number PF9-00065 awarded by the Chandra X-ray Center, which is operated by the Smithsonian Astrophysical Observatory for NASA under contract NAS8-03060. DG acknowledges support from the Lyman Spitzer, Jr. Fellowship awarded by the Department of Astrophysical Sciences at Princeton University.  DG also acknowledges support from the Fermi 4 cycle grant \#041305.  PM acknowledges the support from the European Research Council (grant CAMAP-259276), and the partial support of grants AYA2010-21097-C03-01, CSD2007-00050, and PROMETEO-2009-103. The simulations have been performed on the {\emph{Llu\'{\i}s Vives}} cluster at the University of Valencia.

\bibliographystyle{mn2e}
\bibliography{ms2}



\end{document}